\documentclass[pra,aps,twocolumn,floatfix,showpacs,tightenlines,superscriptaddress,amsmath,amssymb]{revtex4}
\usepackage{amssymb,amsbsy,epsfig,color,graphicx}
\usepackage{graphicx}
\usepackage{dcolumn}
\usepackage{bm}

\newcommand{\ket}[1]{|#1\rangle}
\newcommand{\bra}[1]{\langle#1|}

\newcommand{\virg}[1]{``#1'' }
\newcommand{\eq}[1]{Eq.~(\ref{#1})}

\begin{document}

\title{Determination of ground state properties in quantum spin systems \\
       by single qubit unitary operations and entanglement excitation energies}

\author{S. M. Giampaolo}
\affiliation{Dipartimento di Matematica e Informatica, Universit\`a
degli Studi di Salerno, Via Ponte don Melillo, I-84084 Fisciano
(SA), Italy} \affiliation{CNR-INFM Coherentia, Napoli, Italy, CNISM
Unit\`a di Salerno, and INFN Sezione di Napoli, Gruppo collegato di
Salerno, Baronissi (SA), Italy}

\author{F. Illuminati}
\thanks{Corresponding author. E-mail: illuminati@sa.infn.it.}
\affiliation{Dipartimento di Matematica e Informatica, Universit\`a
degli Studi di Salerno, Via Ponte don Melillo, I-84084 Fisciano
(SA), Italy} \affiliation{CNR-INFM Coherentia, Napoli, Italy, CNISM
Unit\`a di Salerno, and INFN Sezione di Napoli, Gruppo collegato di
Salerno, Baronissi (SA), Italy} \affiliation{ISI Foundation for
Scientific Interchange, Viale Settimio Severo 65, I-00173 Torino,
Italy}

\author{P. Verrucchi}
\affiliation{CNR-INFM SMC, and Dipartimento di Fisica, Universit\`a
di Firenze, via G. Sansone 1, I-50019 Sesto Fiorentino, Italy}

\author{S. De Siena}
\affiliation{Dipartimento di Matematica e Informatica, Universit\`a
degli Studi di Salerno, Via Ponte don Melillo, I-84084 Fisciano
(SA), Italy} \affiliation{CNR-INFM Coherentia, Napoli, Italy, CNISM
Unit\`a di Salerno, and INFN Sezione di Napoli, Gruppo collegato di
Salerno, Baronissi (SA), Italy}

\pacs{03.65.Ca, 03.67.Mn, 73.43.Nq, 75.10.Jm}

\begin{abstract}

We introduce a method for analyzing ground state properties of
quantum many body systems, based on the characterization of separability
and entanglement by single subsystem unitary operations. We apply the
method to the study of the ground state structure of several
interacting spin-$1/2$ models, described by Hamiltonians with
different degrees of symmetry. We show that the approach based on
single qubit unitary operations allows to introduce {\it
``entanglement excitation energies''}, a set of observables that
can characterize ground state properties, including the quantification
of single-site entanglement and the determination of quantum critical points.
The formalism allows to identify the existence and location of
factorization points, and a purely quantum {\it ``transition of
entanglement''} that occurs at the approach of factorization. This
kind of quantum transition is characterized by a diverging ratio of 
excitation energies associated to single-qubit unitary operations.
\end{abstract}

\date{December 13, 2007}

\maketitle

\section{introduction}

With the advent of quantum information theory, entanglement has been
recognized as a fundamental physical resource, on equal footing with
other fundamental resources such as energy and entropy. 
The relationships between these resources are being actively 
investigated, and it has been suggested that the
entanglement properties in many-body systems may be characterized by
suitably defined energy observables \cite{Dowling,GhuneT06}. Establishing
{\it direct} connections between energy observables and entanglement
is of interest because the former could lead to a deeper conceptual
understanding of the latter and be exploited for its experimental
production and manipulation in complex quantum systems. On the other
hand, the study of the role played by entanglement in quantum phase
transitions~\cite{Sachdev00,OsterlohEtal02,OsborneN02} and its
relations with ground state (GS) properties has sparked a rapidly
growing field of research~\cite{ReviewFazio}. As entanglement plays
a fundamental role in quantum information~\cite{ReviewHorodecki},
the question then arises whether it is possible to determine
structural aspects of complex quantum systems using concepts and
techniques of quantum information theory. Relevant open challenges
include the understanding of the relation between GS entanglement,
physical observables, and criticality, as well as the determination
of factorization points at which quantum ground states become
separable \cite{KurmannTM82,RoscildeEtal0405}.

Recently, a formalism of {\em single subsystem unitary operations}
has been introduced~\cite{GeneralAnalysis,GaussianCase}: These
operations are defined as unitary and traceless local
transformations with non degenerate spectrum (as well as Hermitian,
in the case of qubits) that act on a single subsystem, leaving
unaffected the remaining ones. The corresponding approach provides a
tool for studying pure state entanglement for systems of qubits
(spin $1/2$) or qutrits (spin $1$)~\cite{GeneralAnalysis}, and for
Gaussian states of continuous variable systems~\cite{GaussianCase}.
For such systems, a necessary and sufficient condition for pure
state separability is the existence of a unique transformation
(termed {\em invariant}) that leaves the state unchanged. On the
other hand, if a pure state is entangled, the minimum Euclidean
distance in Hilbert space from the state and its image under the
action of single subsystem unitary operations singles out a unique,
{\it extremal} operation that is directly related to the
entanglement properties of the state. Namely, the squared minimum
distance coincides exactly with the linear entropy, and is thus
monotone in the entropy of entanglement of the state. As a
consequence of these results, separability points are determined and
singled out by imposing equality of the extremal and invariant
operations.

The approach based on single subsystem unitary operations allows to
gain informations on the global nature of pure states of a composite
system by looking at how states transform under local operations
that, by definition, do not change the content of entanglement
(local unitaries). Moreover, the method finds a natural operational
interpretation, as looking at the response to a given action is a
basic tool in the investigation of physical properties.

The structure of the paper is as follows: In Sec. \ref{s.SQUOs} we
introduce Single Qubit Unitary Operations (SQUOs), and show that
each SQUO singles out a direction in the three-dimensional
single-spin space; we then define the {\it extremal} SQUO, as the
one associated to the distance of the ground state (GS) from the 
nearest state in the set of states obtained from the GS under the 
action of SQUOs. In Sec. \ref{s.EXE}, we define the excitation energy
relative to each SQUO, and demonstrate that, under rather general
conditions, the excitation energy associated with the extremal SQUO
vanishes if and only if the GS is factorized: We therefore name it
{\it entanglement excitation energy} (EXE). Sec. \ref{s.ISS} is
devoted to the analysis of several $S=1/2$ one-dimensional spin
models, with attention focused on the dependence of the EXE on the
Hamiltonian parameters. In Sec. \ref{s.results} we present and
discuss our results: we find that the EXE follows monotonically the
behavior of the GS single-site entanglement (i.e. the von Neumann
block entropy between a single spin and the rest of the system) and
exhibits signatures of the relevant GS properties, including the
existence and location of factorization and quantum critical
points. Finally, we introduce the {\it orthogonal} SQUOs as those
defined by two directions orthogonal to each other, and to the
direction selected by the extremal SQUO: We find that the
corresponding excitation energies coincide when the EXE vanishes, in
such a way that the ratio between their difference and the EXE
itself diverges at the approach of a factorization point in a large
class of models: such a divergence may therefore be exploited in
order to detect the {\it entanglement transition} associated with
the occurrence of a fully separable ground
state~\cite{AmicoEtal06,BaroniEtal07} in strongly correlated quantum
systems. In Sec.~\ref{s.conclusions} we draw our conclusions and
discuss some possible future lines of research.

\section{single qubit unitary operations}
\label{s.SQUOs}

Consider a $N$-qubit system. A SQUO is
defined~\cite{GeneralAnalysis} as the following unitary
transformation
\begin{equation}
U_{k} \equiv \bigotimes_{i \neq k} \textbf{1}_i \otimes {\cal{O}}_k~,
\label{e.Uk}
\end{equation}
where the operators $\textbf{1}_i$ are the identities on the $N-1$
spins different from spin $k$, and the operator ${\cal{O}}_k$ is a
Hermitian, unitary, and traceless $2\times 2$ transformation, acting
on spin $k$. In the standard basis $\{ \ket{\uparrow},
\ket{\downarrow}\}$ , ${\cal{O}}_k$ is parametrized as
\begin{equation}
{\cal{O}}_k(\theta,\varphi) \, = \, \left(
\begin{array}{cc}
\cos{\theta} & \sin \theta e^{-i \varphi} \\
\sin \theta e^{i \varphi} & - \cos{\theta} \\
\end{array}
\right) \; ,
\label{e.Ok}
\end{equation}
where $\theta$ and $\varphi$ vary in the ranges $[-\pi/2, \pi/2)$
and $[0,2 \pi)$, respectively. The above expression may be written
as ${\cal{O}}_k=u\cdot{\bm{\sigma}}_k$, where
${\bm{\sigma}}_k=(\sigma_k^x,\sigma_k^y,\sigma_k^z)$ are the Pauli
matrices, and the unitary vector
$u\equiv(u_x,u_y,u_z)=(\sin\theta\cos\varphi,\sin\theta\sin\varphi,\cos\theta)$
defines a direction in the three-dimensional single-spin space: each
direction in such a space defines a SQUO and, viceversa, to each
SQUO there corresponds one such direction. The action of a SQUO
transforms any pure state $\ket{\Psi}$ defined in the $2^N$
dimensional Hilbert space in the state $U_{k}\ket{\Psi}$. If
$U_{k}\ket{\Psi}=\ket{\Psi}$ then $U_k$ is the {\em invariant} SQUO.
In general, however, the transformed state differs from
$\ket{\Psi}$: This difference can be quantified by introducing the
standard Hilbert-Schmidt distance:
\begin{eqnarray}
d(U_k;\ket{\Psi})&{\equiv}&\sqrt{1{-}|\bra{\Psi}U_k\ket{\Psi}|^{2}}=\nonumber\\
& & \nonumber \\
&{=}&\sqrt{1{-}|\bra{\Psi}u\cdot{\bm{\sigma_k}}\ket{\Psi}|^2}\equiv 
d(u;\ket{\Psi})~.
\label{e.d}
\end{eqnarray}
Minimizing the distance over the entire set of SQUOs, i.e.
evaluating $\min_{\{\theta,\varphi\}} d(u;\ket{\Psi})$, yields that
the distance of a pure state from the nearest transformed state is obtained for
\cite{GeneralAnalysis}:
\begin{eqnarray}
\widetilde{\varphi} & = & \arctan{\left( \frac{M^y_k}{M^x_k} \right)}
\; , \nonumber \\
& & \nonumber \\
\widetilde{\theta} & = & \arctan{\left( \frac{M^x_k
\cos{\widetilde{\varphi}}
+
M^y_k \sin{\widetilde{\varphi}}}{M^z_k} \right)} \; ,
\label{e.minimumpoint}
\end{eqnarray}
where $M^\alpha_k$ denotes the spin expectation values
$\langle\sigma^\alpha_k\rangle/2$ ($\alpha = x,y,z$) on the state
$\ket{\Psi}$. The values (\ref{e.minimumpoint}) fix the direction
$\widetilde{u}$ which determines the {\it extremal} SQUO,
$U_{k}^{extr}$. The distance $d(\widetilde{u};\ket{\Psi})$
of a state from the nearest transformed state, corresponding to 
the extremal SQUO $U_{k}^{extr}$, can be evaluated 
explicitly, and its square reads
\begin{equation}
d^{2}(\widetilde{u};\ket{\Psi}) = 1 - \left[
(M_{k}^{x})^2 + (M_{k}^{y})^2 + (M_{k}^{z})^2 \right] \; .
\label{e.MinimumSquaredDistance}
\end{equation}
This is exactly the linear entropy $S_L$ of state $\ket{\Psi}$:
$S_{L} = 2(1 - \mathrm{Tr}[\rho^{2}_k])$, where $\rho_k$ is the
single-site reduced density matrix after tracing out the $N-1$
remaining spins. It is a well known fact that in the case of qubits
the linear entropy coincides with the tangle $\tau$: $S_{L} = \tau \equiv
4\mathrm{Det}\rho_k$ \cite{CoffmanKW00,OsborneV}. The pure
state entanglement, measured by the von Neumann entropy ${\cal{E}}$
\cite{Bennet}, is a single-valued, monotonic function of either of
these two coinciding quantities (either the linear entropy or the
tangle): ${\cal{E}}(\ket{\Psi}) = - x{\ln}_{2}x - (1 -
x){\ln}_{2}(1-x)$ where $x = (1 + \sqrt{1 - \tau})/2$. Therefore,
due to \eq{e.MinimumSquaredDistance}, extremal SQUOs and the
associated Euclidean distances determine
pure state entanglement, whose entropic quantification is recast in
direct geometric terms.

\section{Entanglement Excitation Energies}
\label{s.EXE}

The full consequences of the above results can be exploited for the
characterization of GS properties. In the following, we show that
$U_{k}^{extr}$ can be used to construct an energy observable which
witnesses GS separability and quantifies GS entanglement.

Let $\ket{G}$ be the GS of a system of interacting spins with
Hamiltonian ${\cal{H}}$, and define the excitation energies,
\begin{equation}
\Delta E(U_k)
\equiv \bra{G}U^\dagger_k{\cal{H}}U_k\ket{G} - \bra{G} {\cal{H}}\ket{G} ~;
\label{e.EEs}
\end{equation}
the dependence of $\Delta E(U_k)$ on the parameters $\theta$ and
$\varphi$, i.e. on the direction $u$ selected by the SQUO $U_k$, is
made evident by using Eqs.~(\ref{e.Uk}) and (\ref{e.Ok}). 
One obtains that
\begin{eqnarray}
\Delta E(U_k)&=&u^2_x\Delta E(\sigma_k^x)+
u^2_y\Delta E(\sigma_k^y)+u^2_z\Delta E(\sigma_k^z)+ \nonumber\\
& &+ u_xu_y\epsilon_{xy}+ u_xu_z\epsilon_{xz}+ u_yu_z\epsilon_{yz}~,
\label{e.EEs.u}
\end{eqnarray}
where
$\epsilon_{\alpha,\beta}=\epsilon_{\beta,\alpha}\equiv\bra{G}\sigma_k^\alpha{\cal{H}}\sigma_k^\beta+
\sigma_k^\beta{\cal{H}}\sigma_k^\alpha\ket{G}$.

Amongst these excitation energies, which are non-negative by
definition, the one corresponding to the extremal SQUO, $\Delta
E(U_{k}^{extr})$, has a prominent role in the analysis  of
entanglement properties of the GS, due to the intimate relation
between its vanishing and the occurrence of full separability. We
will therefore name it {\it entanglement excitation energy} (EXE).

In fact, if in the GS the spin $k$ is not entangled with the rest of
the system, then $U_k^{extr}\ket{G} =\ket{G}$, and the EXE
consequently vanishes. On the other hand, if the spin $k$ is
entangled with the rest of the system, one necessarily has
$U_k^{extr}\ket{G}\neq \ket{G}$ and the EXE may vanish only if
$U_k^{extr}\ket{G}$ is again a state of minimum energy, i.e. if
$\bra{G}[{\cal{H}},U_k^{extr}]\ket{G}=0$. Therefore, barring 
pathological or trivial cases in which the Hamiltonian does
commute with each SQUO, the vanishing of the EXE occurs only 
in the presence of a factorized GS, proving the statement.

It is worth noticing that the vanishing of the EXE is a necessary, 
but not sufficient, condition for the separability of excited states. 
In fact, considering an entangled excited state $\ket{\psi}$, the action 
of the corresponding extremal SQUO transforms it in a different state that is
not, in general, an eigenstate of the Hamiltonian and can have a
non vanishing projection on any possible eigenstate of the system. The
possibility to populate eigenstates with energy lower than that
associated to $\ket{\psi}$ makes then it possible to obtain a vanishing EXE
even in the presence of an extremal SQUO.

\section{interacting spin systems}
\label{s.ISS}

Equipped with these results, we move on to apply them to the
determination of the GS properties of interacting spin systems. We
consider antiferromagnetic anisotropic Heisenberg-like $XYZ$ model
Hamiltonians:
\begin{equation}
\label{ModelXYZ}
{\cal{H}} = \sum_i S_i^x S_{i+1}^x+ \Delta_y S_i^y
S_{i+1}^y + \Delta_z S_i^z S_{i+1}^z-hS_i^z \; ,
\end{equation}
where $S_i^{\alpha}=\sigma_i^\alpha/2$, the index $i$ runs on the
sites of a one-dimensional lattice, and periodic boundary conditions
are assumed. In \eq{ModelXYZ} $h$ is the reduced (dimensionless)
magnetic field, and $\Delta_y$, $\Delta_z$ are the anisotropy
parameters, $0 \le \Delta_{y,z} \le 1$. This choice is motivated by
its great generality: according to the different values of the
anisotropies, \eq{ModelXYZ} comprises many different models
belonging to different classes of universality, including the Ising
and the Heisenberg ones. For the models described by \eq{ModelXYZ}
genuine quantum phase transitions, with or without order parameter,
occur at zero temperature at critical values $h_{\rm c}$ that depend
on the anisotropies. If $\Delta_y=1$, the symmetry forces a
transition without order parameter, and the system remains in a
critical regime for all $h \le h_{\rm c}$. In this case if
$\Delta_z\neq1$ the system is in the $XXZ$ symmetry class and goes
over to the Heisenberg symmetry class when $\Delta_z=1$. For $0 \le
\Delta_y <1$, the system undergoes a second order phase transition,
acquiring a non-vanishing staggered magnetization along the $x$
direction for $h <h_{\rm c}$. Antiferromagnetic order divides the
lattice in two sublattices, each characterized by opposite value of
the magnetization. Again, changing the value of $\Delta_z$ lets the
system move from the Ising symmetry class $(\Delta_z=0)$ to the
$XYZ$ symmetry class $(\Delta_z \neq 0)$. Concerning the
magnetization along the directions $y$ and $z$, one has that,
regardless of the anisotropy, the magnetization $M_y$ is always
vanishing, while $M_z$ vanishes at and only at $h = 0$. Importantly,
in the models described by \eq{ModelXYZ} the external field not only
drives the system through a quantum phase transition, but induces as
well the factorization of the GS, for $h=h_{\rm f}\equiv
\sqrt{(1+\Delta_z)(\Delta_y+\Delta_z)}$ \cite{KurmannTM82}. This
phenomenon has been recently investigated in relation to the
analysis of entanglement properties in spin systems
\cite{RoscildeEtal0405}. For the ground states of the models
described by \eq{ModelXYZ}, the vanishing of the EXE is a necessary
and sufficient condition for GS separability.

Using Eqs.~(\ref{e.minimumpoint}), and considering that it is always
$M_k^y=0$, and hence ${\widetilde{\varphi}}=0$, on gets, by
Eq.~(\ref{e.EEs.u}), the exact expression of the EXE that reads
\begin{eqnarray}
\label{e.DeltaEtilde}
\Delta E(\widetilde{u})&{\equiv}&\Delta E(U^{extr}) = \nonumber \\
&{=}&-16 \left[ \frac{g_{xx} M_z^2-M_x M_z g_{zx}}{1-\tau}
+ \frac{\Delta_y g_{yy}}{4} \right. \nonumber \\
& & \left. + \frac{ \Delta_z \left(g_{zz} M_x^2 + M_x M_z
g_{zx}\right)}{1-\tau}\right] \; ,
\end{eqnarray}
where $\tau = S_L$ is the GS tangle (linear entropy), $g_{\alpha
\beta}=\bra{G}S_i^{\alpha}S_{i+1}^{\beta}\ket{G}$ are the
nearest-neighbor correlation functions,
$M_\alpha=\bra{G}S_i^{\alpha}\ket{G}$ are the expectation values of
the spin operators, and any dependence on the site index is neglected due
to translational invariance.

\begin{figure}[t]

\includegraphics[width=7.5cm]{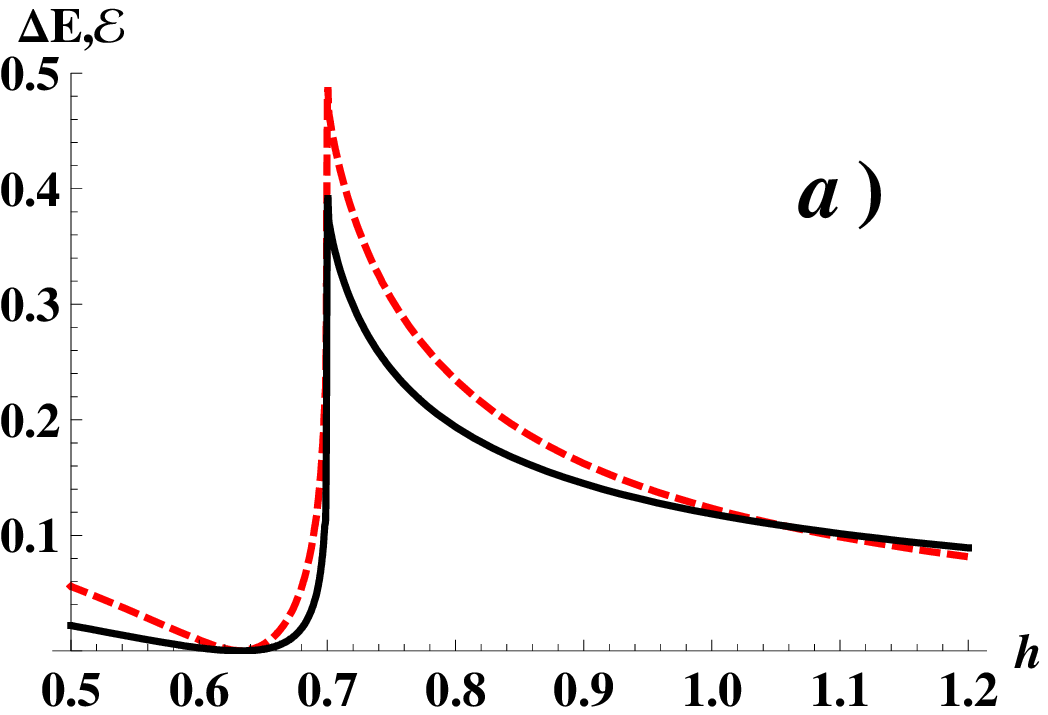}
\vskip .5truecm
\includegraphics[width=7.5cm]{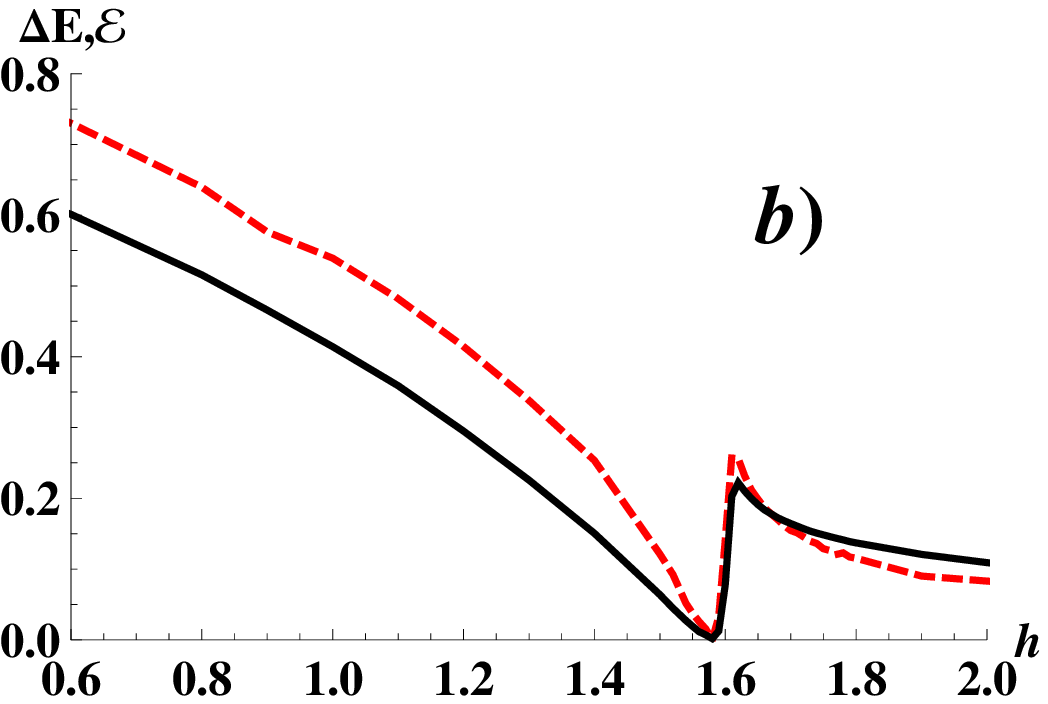}

\caption{(color online) Entanglement excitation energy $\Delta
E(\widetilde{u})$ (Black solid line) and single-site entanglement
${\cal{E}}$ (Red dashed line) as functions of the reduced field $h$,
for {\it a)}: $\Delta_y = 0.4$, $\Delta_z = 0$ ($XY$), and {\it b)}:
$\Delta_y = 0.25$, $\Delta_z = 1$ ($XYZ$). All quantities are
dimensionless.} \label{f.EXE-S.XY}
\end{figure}

\begin{figure}[t]

\includegraphics[width=7.5cm]{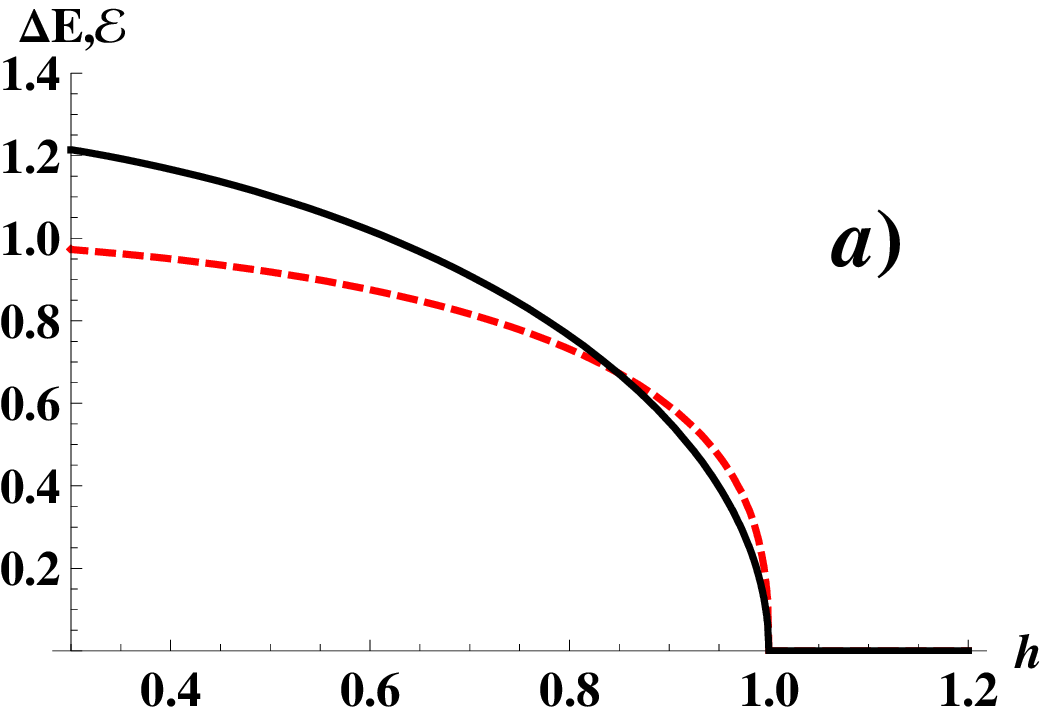}
\vskip .5truecm
\includegraphics[width=7.5cm]{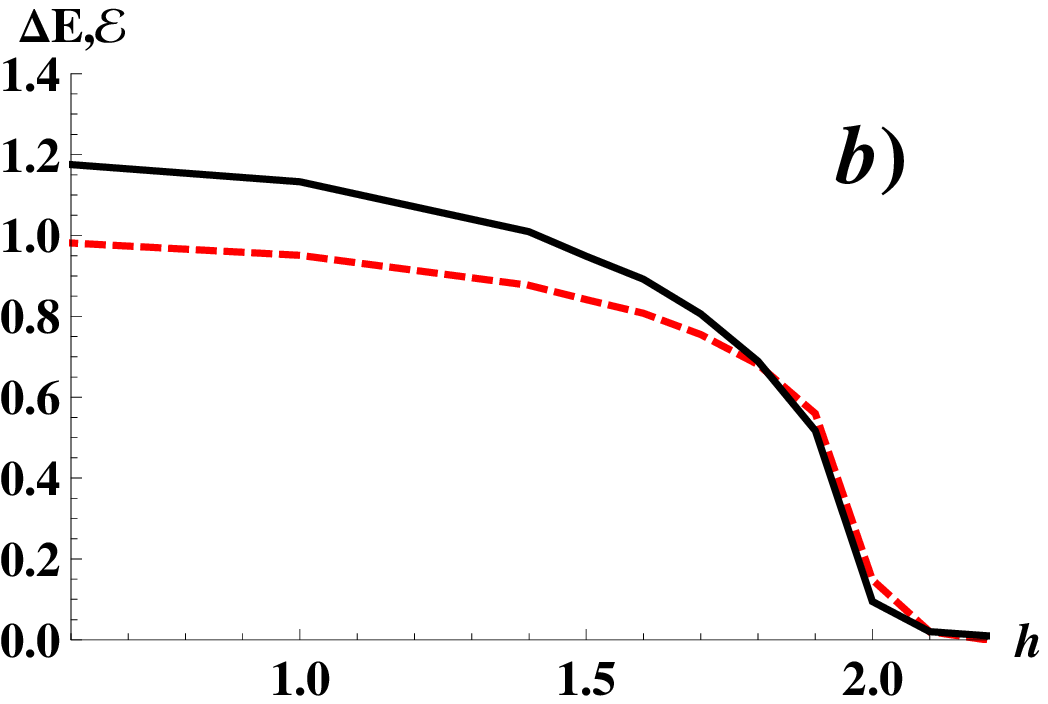}

\caption{(color online) Entanglement excitation energy $\Delta
E(\widetilde{u})$ (Black solid line) and single-site entanglement
${\cal{E}}$ (Red dashed line) as functions of the reduced field $h$,
for {\it a)}: $\Delta_y = 1$, $\Delta_z = 0$ ($XXZ$), and {\it b)}:
$\Delta_y = 1$, $\Delta_z = 1$ (Heisenberg). All quantities are
dimensionless.} \label{f.EXE-S.XX}
\end{figure}

\begin{figure}[t]

\includegraphics[width=7.5cm]{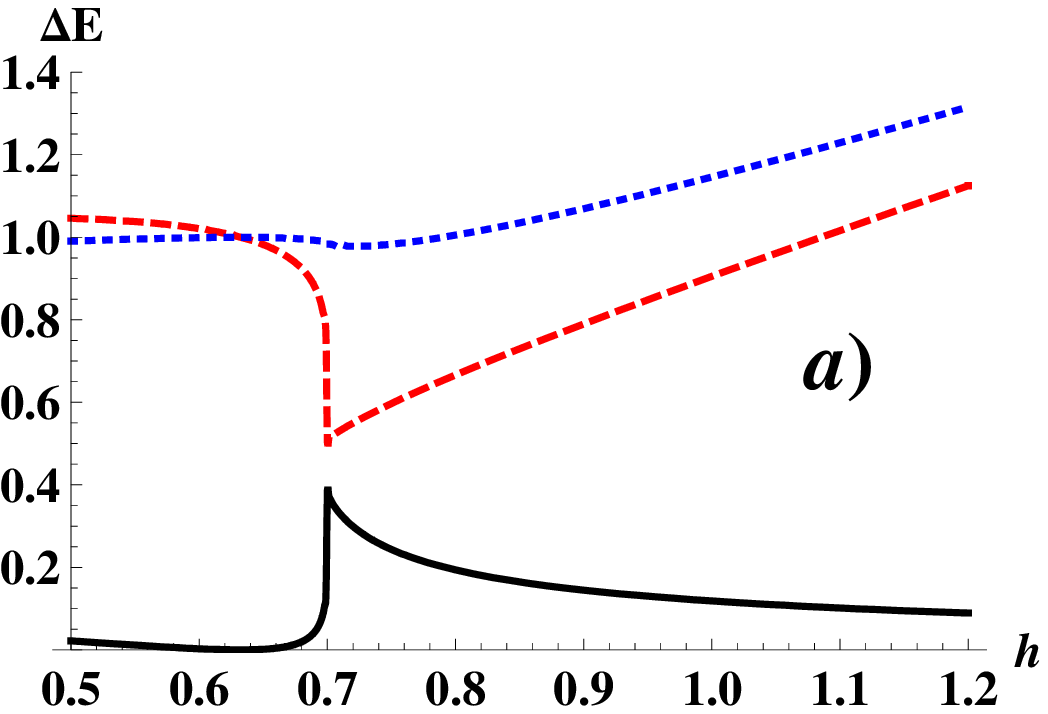}
\vskip .5truecm
\includegraphics[width=7.5cm]{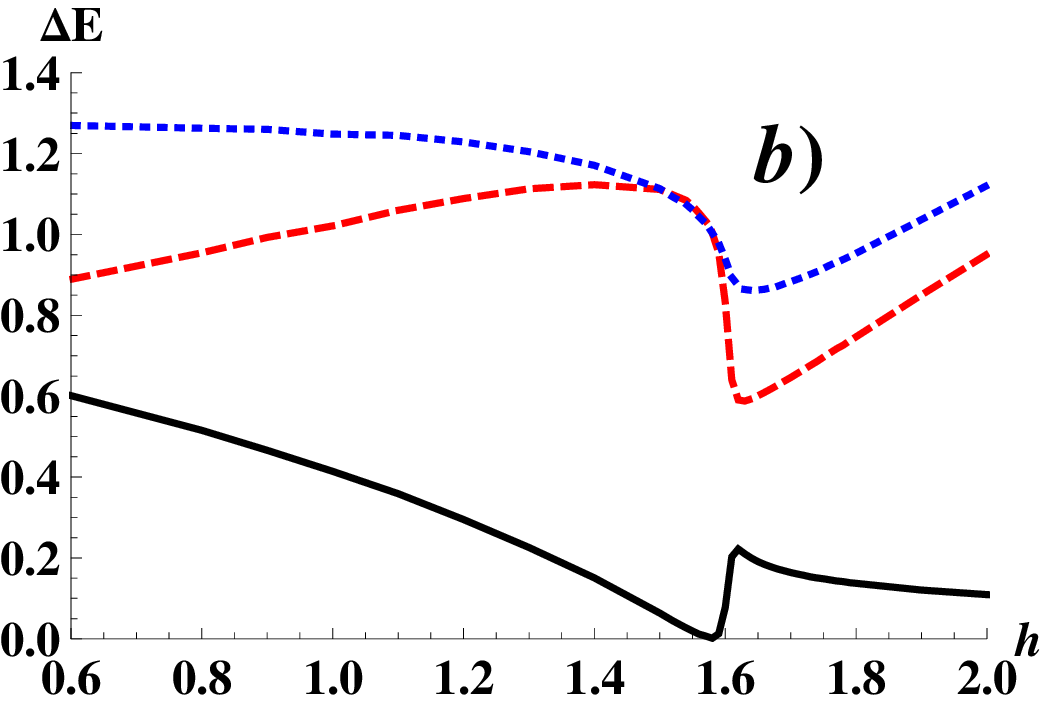}

\caption{(color online) Entanglement excitation energy
$\Delta E(\widetilde{u})$ (Black solid
line), excitation energy $\Delta E'_\perp$ (Red dashed line), and 
excitation energy $\Delta E''_\perp$
(Blue dotted line) as functions of $h$, for {\it a)}: $\Delta_y = 0.4$,
$\Delta_z = 0$ ($XY$), and {\it b)}: $\Delta_y = 0.25$, $\Delta_z =
1$ ($XYZ$). All quantities are dimensionless.} \label{f.DeltaE.all}
\end{figure}

\begin{figure}[t]

\includegraphics[width=7.5cm]{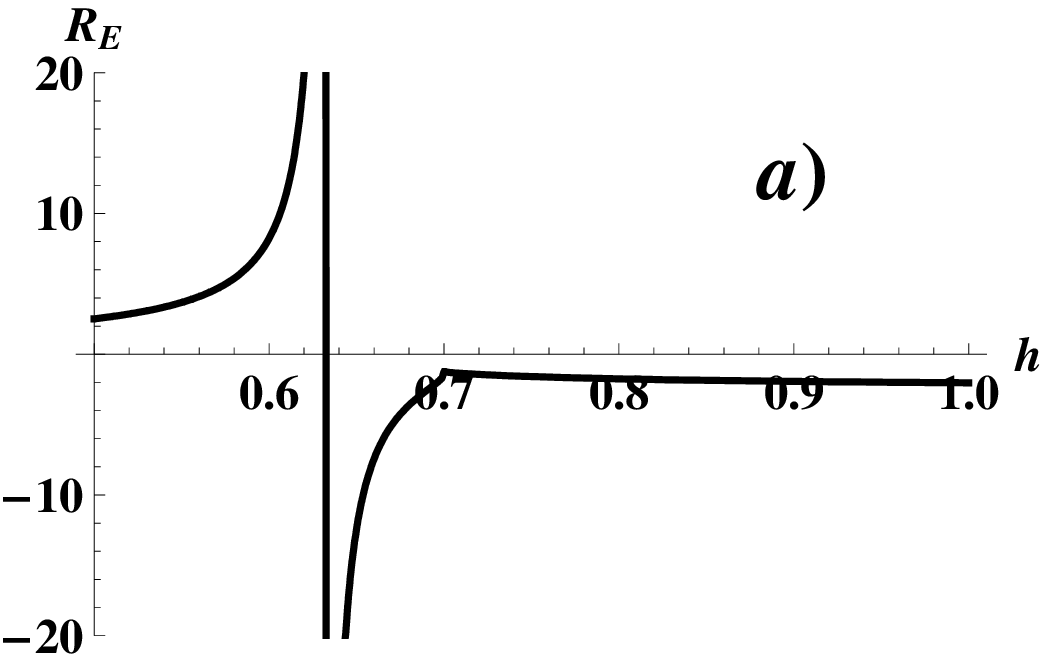}
\vskip .5truecm
\includegraphics[width=7.5cm]{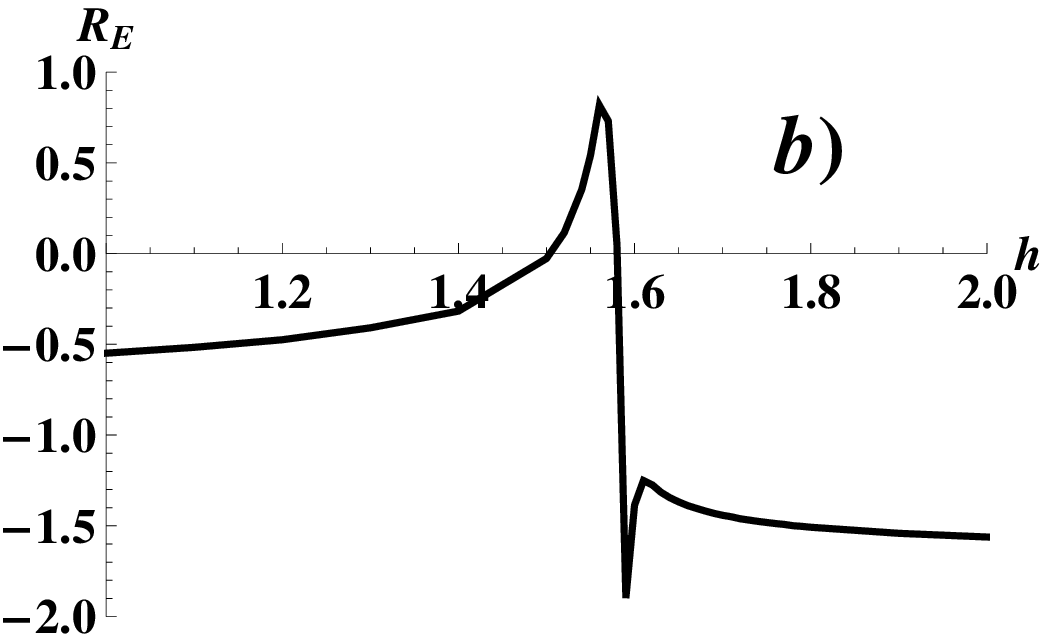}

\caption{(color online) Entanglement energy ratio $R_{E}$ as a function of $h$,
for {\it a)}: $\Delta_y = 0.4$, $\Delta_z = 0$ ($XY$), and {\it b)}:
$\Delta_y = 0.25$, $\Delta_z = 1$ ($XYZ$).  All quantities are
dimensionless.} \label{f.Re}
\end{figure}

\section{Results}
\label{s.results}

We first analyze the EXE, $\Delta E(\widetilde{u})$, and compare it
with the von Neumann entropy ${\cal{E}}$. The latter measures the
bipartite single-site entanglement between one selected spin and the
rest of the chain \cite{Bennet}, and it provides an upper bound to
all bipartite block entanglements. Therefore, its vanishing
guarantees the full separability of the GS. Exploiting the
conditions for the vanishing of $S_L$ and $\Delta E(\widetilde{u})$
allows to determine unambiguously the value and location of the
factorization point.

In Figs.~\ref{f.EXE-S.XY} and \ref{f.EXE-S.XX} we plot $\Delta
E(\widetilde{u})$ and ${\cal{E}}$ as functions of $h$.
Fig.~\ref{f.EXE-S.XY} shows exact analytical
results~\cite{Barouch,Pfeuty70} for the XY model, and Quantum Monte
Carlo (QMC) data for the XYZ model. Fig.~\ref{f.EXE-S.XX} shows
exact analytical results~\cite{Barouch} for the XX model, and QMC
data for the Heisenberg model. Numerical data have been obtained via
stochastic-series-expansions quantum Monte Carlo simulations
\cite{SyljuasenEtal02,RoscildeEtal0405}, based on a modified
directed-loop algorithm. The EXE and the entropy of entanglement
evidently exhibit the same behavior, vanishing simultaneously at the
factorization points. At critical points, the derivatives of both
$\Delta E(\widetilde{u})$ and ${\cal{E}}$, with respect to $h$,
diverge, signaling the onset of a quantum phase transition. This
divergence occurs both for $\Delta_z=1$ and $\Delta_z<1$, at
variance with the concurrence, i.e. a measures of the entanglement
between two spins of the lattice \cite{Wootters98}, that exhibit
different behaviors at the critical points for models with different
symmetries \cite{ReviewFazio}. In this sense, we suggest that the 
EXE, contrary to other observable estimators of entanglement, can be 
considered a {\it universal} indicator for the onset of quantum phase
transitions in spin systems.

The relevance of SQUOs and EXEs in determining entanglement and
factorization properties, and in general qualitative changes in the
GS can be understood in physical terms and related to ``entanglement
transitions''\cite{AmicoEtal06,BaroniEtal07}. If a system is in a
pure classical state the orientation of any spin is well defined.
Selecting a spin and performing a rotation about its orientation
leaves the state of the system and its energy unchanged. Viceversa,
if the system is in an entangled quantum state the single-spin
orientation is not defined and any \virg{rotation}, i.e. any SQUO,
will change the state of the system and will yield a change in
energy. Therefore, the increase in energy is connected to \virg{how
much the GS is entangled}. Constructing the excitation energy
associated to the extremal SQUO, i.e. the EXE, formalizes the
argument.

Based on the above comparison with the classical case, let us
introduce two directions, $u'_\perp$ and $u''_\perp$,
orthogonal to each other and to the direction $\widetilde{u}$
associated to the extremal SQUO $U^{extr}$. Being
$\widetilde{u}=(\sin\widetilde{\theta},0,\cos\widetilde{\theta})$,
we choose $u'_\perp=(\cos\widetilde\theta,0,-\sin\widetilde\theta)$
and $u''_\perp=(0,1,0)$. These directions define the {\em
orthogonal} SQUOs, via Eq.~(\ref{e.Ok}), and the corresponding
excitation energies $\Delta E'_\perp$ and $\Delta E''_\perp$, via
Eq.~(\ref{e.EEs.u}). In Fig.~\ref{f.DeltaE.all} we compare the field
dependence of the EXE with that of the above defined excitation
energies, for the $XY$ and $XYZ$ models: We find that $\Delta
E'_\perp$ and $\Delta E''_\perp$ coincide at $h_{\rm f}$, i.e. when
the GS is factorized, in full analogy with the classical case. In
fact if the system is in a pure classical state any rotation of
$\pi/2$ around a direction orthogonal to the spin orientation causes
an increase in energy by one (in units of the reduced field). For $h
\neq h_f$, $\Delta E'_\perp$ and $\Delta E''_\perp$ behave
differently, a fact which has no classical analogue, so that the
deviation from classicality can be quantified by the difference
$\Delta E'_\perp- \Delta E''_\perp$.

We can then compare such difference to the amount of
entanglement, as measured by the EXE, by defining the 
{\it entanglement energy ratio} (EER)
\begin{equation}
\label{ratio}
R_{E} = \frac{\Delta E'_\perp-\Delta E''_\perp}
{\Delta E(\widetilde{u})} \; .
\end{equation}
In Fig.~\ref{f.Re} we show the behavior of the EER $R_{E}$,
as a function of $h$, for two models belonging, respectively,
to the $XY$ and to the $XYZ$ universality class. In both cases, $R_{E}$
diverges at the factorization point $h_f$, and its first derivative
diverges at the critical point $h_c$. The abrupt change in the GS
properties at the factorization point, as measured by the divergence
of the EER, signals a qualitative change of purely quantum nature, a
``transition of entanglement'' that occurs when $h_f$ is approached.
We remark that all the conventional on-site expectations and $n$-point 
correlation functions remain finite and analytic at a factorization point,
confirming that the divergence of the EER cannot be associated to any
kind of conventional quantum phase transition. On the other hand, the
EER remains fixed at the constant value $R_{E} = -1$ if we consider models of
interacting spins with $\Delta_y=0$: The fact that in this case the EER 
fails to identify the transition from entanglement to factorizability 
is intriguing and deserves further studies.

\section{conclusions and outlook} 
\label{s.conclusions}

In summary, we have defined entanglement excitation energies (EXEs)
associated to the extremal single qubit unitary operations (SQUOs). We have
showed that EXEs are useful tools for the study of various 
ground state properties for interacting spin systems, including
the determination of factorizability, the quantification of single-site 
entanglement, and the identification of quantum critical
points. We have discussed how SQUOs and EXEs determine unambiguously
the presence or the absence of a separable GS, and we have introduced
an entanglement energy ratio (EER) of excitation energies that diverges 
(for $\Delta_y\neq 0$) at the approach of a factorization point, thus
defining an entanglement-separability transition of purely quantum origin.

Considering possible future developments along this line of research,
we would like to observe that when moving away from the factorization point, 
the physics of quantum ground states can in principle be recovered by 
expanding the single-site entanglement in powers of the EXE, as the latter 
naturally defines a small expansion parameter of the theory around
a factorization point. The analytic generalization of the formalism to 
general spin models is a subject of current investigation. The results of such an analysis
could allow to determine exact solutions of non exactly solvable models at 
factorization points, as well as the approximate description of GS properties
out of the factorization point by controlled expansions in powers of the EXEs.
The necessary and sufficient conditions for factorizability obtained within
the formalism of SQUOs might be extended to the case of models involving 
interacting systems of arbitrary local dimension. In particular, it would
be very interesting to consider interacting systems of very high spins and
study them in a continuous-variable representation, for which single-subsystem
unitary operations can be readily expressed in terms of single-mode unitary
transformations. This line of investigation could then be extended as well
to include models of interacting harmonic and anharmonic chains and lattices.

Conceptually, the present work 
discloses an intimate connection between two different universal
physical resources, energy and entanglement, and might have
practical consequences for the experimental production and
manipulation of entanglement, and information transfer in real
systems of interacting qubits. The single-qubit excitation energy
provides a well-defined form of macroscopic observable for the
detection and determination of single-site entanglement in many-body
systems, along lines close in spirit to pair-entanglement detection
in systems of interacting magnetic dipoles by measurements of heat
capacities and magnetic susceptibilities \cite{Ghosh}. As a working
example, the method has been applied and illustrated in quantum
spin-$1/2$ models, that are of particular relevance both for quantum
information and condensed matter physics. However, in principle it
can be applied as well to more general instances
\cite{GeneralAnalysis,GaussianCase}, such as systems of interacting
qutrits, Hubbard models, and harmonic lattices of continuous
variables. The choice of nearest-neighbor couplings in the case of
interacting qubits deserves a comment. In fact, the method is in no
way limited by the choice of the interaction. Generalizations to
spin models with interactions of arbitrary range and lattices of
different topologies are possible and, as suggested above, can be of particular
relevance in establishing the existence of factorization points and
in constructing consistent descriptions of GS quantum physics by
systematic expansions in powers of the EXE around the factorized
solutions.

\section{aknowledgements}

We are grateful to Tommaso Roscilde for providing us with relevant
Quantum Monte Carlo data. SMG and FI wish to thank Rosario Fazio for
helpful discussions. This work was supported by MIUR under the
2005-2007 PRIN-COFIN National Project Program, CNR-INFM Coherentia,
ISI Foundation, and INFN.


\begin{thebibliography}{99}

\bibitem{Dowling} M. R. Dowling, A. C. Doherty, and S. D. Bartlett,
Phys. Rev. A {\bf 70}, 062113 (2004). 

\bibitem{GhuneT06} O. G\"uhne and G. Toth, Phys. Rev. A {\bf 73}, 052319 (2006).

\bibitem{Sachdev00} S. Sachdev, {\em Quantum Phase Transitions} (Cambridge University
Press, Cambridge, 2000).

\bibitem{OsterlohEtal02} A.~Osterloh, L.~Amico, G.~Falci, and R.~Fazio, Nature {\bf 416}, 608
(2002).

\bibitem{OsborneN02}
T.~J.~Osborne and M.~A.~Nielsen, Phys. Rev. A {\bf 66}, 032110 (2002).

\bibitem{ReviewFazio} For a comprehensive review, see
L. Amico, R. Fazio, A. Osterloh, and V. Vedral, arXiv:quant-ph/0703044,
and references therein.

\bibitem{ReviewHorodecki} For a recent overview, see
R. Horodecki, P. Horodecki, M. Horodecki, and K. Horodecki, arXiv:quant-ph/0702225,
and reference therein.

\bibitem{KurmannTM82} J. Kurmann, H. Thomas, and G. M\"uller, Physica A
(Amsterdam) {\bf 112}, 235 (1982).

\bibitem{RoscildeEtal0405} T. Roscilde, P. Verrucchi, A. Fubini, S. Haas,
and V. Tognetti, Phys. Rev. Lett. {\bf 93}, 167203 (2004); {\it
ibidem} {\bf 94}, 147208 (2005).

\bibitem{GeneralAnalysis} S. M. Giampaolo and F. Illuminati, Phys. Rev. A {\bf 76}, 042301 (2007).

\bibitem{GaussianCase} G. Adesso, S. M. Giampaolo, and F. Illuminati, Phys. Rev. A {\bf 76}, 042334 (2007).

\bibitem{AmicoEtal06}
L.~Amico, F.~Baroni, A.~Fubini, D.~Patan\`e, V.~Tognetti, and
P.~Verrucchi, Phys. Rev. A {\bf 74}, 022322 (2006).

\bibitem{BaroniEtal07} F.~Baroni, A.~Fubini, V.~Tognetti, and P.~Verrucchi,  J. Phys. A:
Math. Theor. {\bf 40}, 9845 (2007).

\bibitem{CoffmanKW00} V.~Coffman, J.~Kundu, and W.~K.~Wootters, 
Phys. Rev. A {\bf 61}, 052306 (2000).

\bibitem{OsborneV} T.~J.~Osborne and F.~Verstraete, 
Phys. Rev. Lett. {\bf 96}, 220503 (2006).

\bibitem{Bennet} C. H. Bennett, H. J. Bernstein, S. Popescu, and B.
Schumacher, Phys. Rev. A {\bf 53}, 2046 (1996).

\bibitem{Barouch} E. Barouch, B. McCoy, and M. Dresden,
Phys. Rev. A {\bf 2}, 1075 (1970); E. Barouch and B. McCoy,
Phys. Rev. A {\bf 3}, 786 (1971).

\bibitem{Pfeuty70} P. Pfeuty, Ann. Phys. (N.Y.) {\bf 57}, 79 (1970).

\bibitem{SyljuasenEtal02} O.~F.~Syljuasen and A. W. Sandvik, Phys. Rev. E {\bf 66}, 046701 (2002).

\bibitem{Wootters98} W.~K.~Wootters, Phys. Rev. Lett. {\bf 80}, 2245 (1998).

\bibitem{Ghosh} S. Ghosh, T. Rosenbaum, G. Aeppli, and S. Coppersmith,
Nature {\bf 425}, 48 (2003).

\end{thebibliography}
\end{document}